\documentclass[aps,prb,twocolumn,showpacs]{revtex4}
\usepackage{amsmath}
\usepackage{graphicx,epsfig,psfrag}
\usepackage{amssymb}
\newcommand{\nn}{\nonumber}

\newcommand{\bk}{\mathbf{k}}

\newcommand{\br}{\mathbf{r}}

\begin{document}
\title{Sign change of the Gr\"uneisen parameter and 
magnetocaloric effect near quantum critical points}
\author{Markus Garst$^{(a)}$ and Achim Rosch$^{(b)}$}
\affiliation{$^{(a)}$ Theoretical Physics Institute, University of Minnesota,
  Minneapolis, Minnesota 55455, USA
\\
$^{(b)}$ Institute for Theoretical Physics, University of Cologne, 50937
Cologne, Germany.}
\date{\today} 
\begin{abstract} 
  We consider the Gr\"uneisen parameter and the magnetocaloric effect near a
  pressure and magnetic field controlled quantum critical point, respectively.
  Generically, the Gr\"uneisen parameter (and the thermal expansion) displays
  a characteristic sign change close to the quantum-critical point signaling
  an accumulation of entropy. If the quantum critical point is the endpoint of
  a line of finite temperature phase transitions, $T_c \propto (p_c-p)^\Psi$,
  then we obtain for $p<p_c$, (1) a characteristic increase $\Gamma \sim
  T^{-{1/(\nu z)}}$ of the Gr\"uneisen parameter $\Gamma$ for $T>T_c$, (2) a
  sign change in the Ginzburg regime of the classical transition, (3) possibly
  a peak at $T_c$, (4) a second increase $\Gamma \sim -T^{-{1/(\nu z)}}$ below
  $T_c$ for systems above the upper critical dimension and (5) a saturation of
  $\Gamma \propto 1/(p_c-p)$.  We argue that due to the characteristic
  divergencies and sign changes the thermal expansion, the Gr\"uneisen
  parameter and magnetocaloric effect are excellent tools to detect and
  identify putative quantum critical points.
\end{abstract}
\maketitle

\section{Introduction}

The competition between two different ground states at a quantum phase
transition leads to novel behavior in thermodynamics as well as in transport.
A prominent example are the heavy fermion compounds whose non-Fermi liquid
behavior is attributed to the presence of a quantum critical point (QCP)
associated with a magnetic instability.

We recently pointed out~\cite{PRL} that the Gr\"uneisen parameter, $\Gamma$,
is an important tool to identify and classify a QCP since it necessarily
diverges near a pressure-driven quantum phase transition with characteristic
exponents\cite{Kuechler03}. In this article we focus on another aspect, i.e.,
the sign change of the Gr\"uneisen parameter. We will argue that generically
the sign of $\Gamma$ (and therefore the thermal expansion) changes as entropy
is accumulated near a quantum critical point.  The sign change in combination
with the divergence leads to strong signatures of the Gr\"uneisen parameter
near a QCP.

A quantum phase transition (QPT) occurs at zero temperature upon tuning an
external parameter like doping, pressure, electric field, magnetic field,
etc.~to a critical value. The underlying QPT manifests itself at finite
temperatures in an unusual sensitivity of thermodynamics on these tuning
parameters. In the following we will focus on QCPs which are tuned with
pressure, $p$ and/or magnetic field, $H$. Generalizations are
straightforward. At $T=0$ the distance to the QCP is determined by the control
parameter which depends on pressure and/or magnetic field, $r=r(p,H)$.  Near
the QCP the control parameter can be linearized around the critical pressure
or the critical magnetic field,
\begin{align} \label{ControlParameter}
r(p,H) = (p-p_c)/p_0 = (H-H_c)/H_0\,
\end{align}
where $p_0$ and $H_0$ are a constant pressure and magnetic field scale,
respectively.  Generically, the critical magnetic field will depend smoothly
on pressure, $H_c = H_c(p)$, and vice versa. Note that a linearization in
magnetic field is only possible if the critical field is large, $H/H_c \ll 1$;
in particular, this is not fulfilled in the case of a zero-field QCP.

The critical contribution to the free energy density is a function of this
control parameter and temperature, $f=f(r,T)$. The sensitivity on the tuning
fields is thermodynamically measured by the derivatives of the free energy
density with respect to $r$. For a pressure-tuned QCP an example is provided
by the thermal expansion,
\begin{align}
\alpha &= \frac{1}{V} \left.\frac{\partial V}{\partial T}\right|_{p, H} = 
- \frac{1}{V_m} \left.\frac{\partial S}{\partial p}\right|_{T, H}
= \frac{1}{V_m} \frac{\partial^2 f(r,T)}{\partial p \partial T}\,,
\end{align}
where $V_m$ is the molar volume. With Eq.~(\ref{ControlParameter}) follows
that the thermal expansion is directly proportional to the mixed derivative
$\partial^2f/(\partial r \partial T)$. When the QPT can be tuned by magnetic
field the same derivative can be accessed by the thermodynamic quantity
$(dM/dT)_{H}$. If the QCP is sensitive to both, $p$ and $H$, the critical part
of the thermal expansion and of the temperature derivative of the
magnetization are expected to be proportional to each other near the QCP,
\begin{align} \label{Relation-PM-Variation}
\frac{V_m \alpha}{(dM/dT)_{H}} = \left.\frac{dH_c}{dp}\right|_{p=p_c}\,.
\end{align}
Their ratio gives the dependence of the critical magnetic field on
pressure. Similar relationships hold for the magnetostriction, compressibility
and differential susceptibility which all yield $\partial^2 f/\partial r^2$.

The main quantity of our interest is the Gr\"uneisen parameter which is
measured by the ratio of thermal expansion and molar specific heat $C_p =
\left.T \frac{\partial S}{\partial T}\right|_p$,
\begin{equation} \label{Grueneisen}
\Gamma = \frac{\alpha}{C_p}
=-\frac{(\partial S/\partial p)_T}{V_m T (\partial S/\partial T)_p}\,.
\end{equation}
Note that we define the Gr\"uneisen parameter with the specific heat at
constant {\em pressure}\cite{Grueneisen,Landau} and not at constant volume (as
often used in the literature). It is the specific heat at constant pressure
which is measured in experiments on pressure-controlled QCPs.  The
corresponding quantity for magnetic field tuning is given by $\Gamma_H = -
(\partial M/ \partial T)_H / C_H$, where $C_H = \left.T \frac{\partial
S}{\partial T}\right|_H$. It will become clear below that $\Gamma_H$ describes
the magnetocaloric effect.

\begin{figure}
\centering
\psfrag{Tsimr}{$T \sim |r|^{\nu z}$}
\includegraphics[width=0.9 \linewidth]{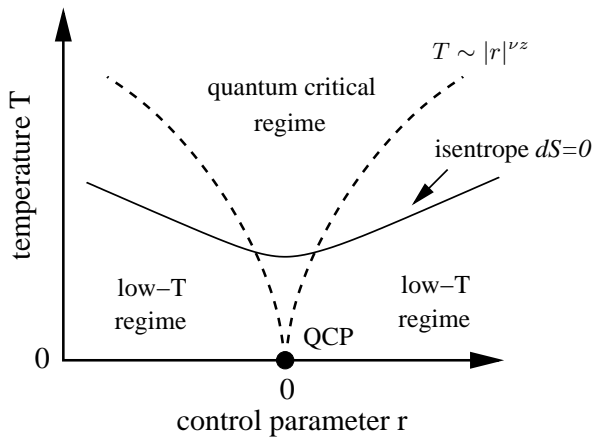}
\caption{\label{Fig:Plane} Different regimes in the phase diagram of a quantum
phase transition. The dotted lines correspond to crossovers between the low-T
and the quantum critical regime, $T \sim |r|^{\nu z}$. The control parameter
might be sensitive to pressure and/or magnetic field. The solid line shows a
generic isentrope along which the entropy is constant, $dS=0$.  }
\end{figure}

Let us shortly recapitulate the main results of Ref.~\onlinecite{PRL}.  The
main observation is that $\Gamma$ diverges at a quantum-critical point, while
it is finite in all non-critical systems or close to generic classical
critical points.  This can most easily be seen in cases where scaling applies
(i.e. for systems below the upper critical dimension, $d+z<4$) and where the
qualitative behavior of $\Gamma$ can be extracted from the scaling form of the
free energy
\begin{align} \label{ScalingForm}
f(r,T) = b^{- (d+z)} f(r\, b^{1/\nu},T\, b^z)\,,
\end{align}
where $b$ is an arbitrary scaling parameter, $d$ is the dimensionality and
$\nu$ and $z$ are the correlation length and dynamical critical exponent,
respectively.  As can be read off directly from Eq.~(\ref{Grueneisen}),
$\Gamma$ scales like $1/r$ or equivalently,
\begin{equation} \label{ScalingDimension}
{\rm dim}[\Gamma_{\rm cr}] = - {\rm dim}[r] = - 1/\nu\,.
\end{equation}
Accordingly,  one obtains directly from (\ref{ScalingForm})
\begin{equation}\label{QCRegime}
\Gamma_{\rm cr} \propto \frac{1}{T^{1/{\nu z}}}\,.
\end{equation}
in the quantum-critical regime, i.e. for $T\gg |r|^{\nu z}$ (see
Fig.~\ref{Fig:Plane}).
On the other hand, in the two low-temperature regimes on the right and left
hand side of the QCP in Fig.~\ref{Fig:Plane}, the Gr\"uneisen parameter
diverges with the inverse of the control parameter $r \propto p-p_c$,
\begin{equation} \label{LowTRegime}
\Gamma_{\rm cr} = - G_r \frac{1}{V_m (p-p_c)}\,.
\end{equation}
Surprisingly, due to the third law of thermodynamics, i.e.~by assuming a
vanishing residual entropy at zero temperature, it is possible to determine
even the prefactor $G_r$ of the divergence from a scaling analysis. It is
given by a simple combination of critical exponents
\begin{equation} \label{UniPrefactor}
G_r = - \nu \frac{y^{\pm}_0 z-d}{y^{\pm}_0}\,,
\end{equation}
where the exponents $y^{+}_0$ and $y^-_0$ are associated with the
low-temperature behavior of the specific heat, $C_p \sim T^{y^\pm_0}$, on the
right and left hand side of the QCP, respectively. As was shown in
Ref.~\onlinecite{PRL}, these results might even hold (up to possible
logarithmic corrections) in situations where the simple scaling Ansatz
(\ref{ScalingForm}) fails, i.e. for systems above the upper critical
dimension.

Equation (\ref{LowTRegime}) implies not only a divergence of $\Gamma$ but also
a sign change (assuming that $G_r$ has the same sign on both sides of the
QCP)! Obviously the question arises where and how this drastic sign change
takes place in the finite-temperature phase diagram. This will be one of the
main topics discussed in this paper.

The following section will discuss the sign changes using qualitative
arguments.  Sec.~\ref{Sec:IsingChain} investigates quantum critical points
where there is no phase transition at finite temperature and briefly discusses
experiments close to metamagnetic quantum phase transitions.  In
Sec.~\ref{Sec:BEC} we study how thermal expansion, Gr\"uneisen parameter and
magnetocaloric effects are influenced by a phase-transition at finite $T$ in
proximity to a QCP. An overview of our main results is given in
Sec.~\ref{Sec:Summary}.

\section{Sign of the Gr\"uneisen parameter}

In order to obtain insight into the meaning of the sign of the Gr\"uneisen
parameter it proves useful to consider a line of constant entropy 
within the pressure-temperature plane $(p,T)$,
\begin{align} 
d S &= 
\left.\frac{\partial S}{\partial T}\right|_{p} dT + 
\left.\frac{\partial S}{\partial p}\right|_{T} dp 
\overset{!}{=} 0\,.
\end{align}
Using the definition of the thermal expansion and the specific heat
we obtain for $\Gamma$, 
\begin{align} \label{PressureCaloric}
\Gamma = \frac{1}{V_m T} \left. \frac{dT}{dp}\right|_S\,.
\end{align}
The Gr\"uneisen parameter measures the variation of temperature upon pressure
changes under constant entropy conditions. The Gr\"uneisen parameter thus
corresponds to a pressure-caloric effect. As already alluded to, for a QPT
that can be driven by magnetic field the quantity analogous to the Gr\"uneisen
parameter is the magnetocaloric effect
\begin{align} \label{MagnetoCaloric}
\Gamma_H = - \frac{(dM/dT)_H}{C_H} = 
\frac{1}{T} \left. \frac{dT}{dH}\right|_S\,,
\end{align}
where $C_H$ is the specific heat at constant $H$.  Experimentally, the
quantities $\Gamma$ and $\Gamma_H$ can be directly accessed by measuring the
change in temperature at constant entropy upon pressure and magnetic
field variations, respectively. In mathematical terms both yield the slope of
the constant entropy curves, i.e.~isentropes in the phase diagram.

How do the isentropes look like near a quantum phase transition?  We expect
that we have an accumulation of entropy near the quantum critical point since
directly at the QCP the system is frustrated of two different possible ground
states. From this expectation follows that the isentropes are tilted towards
the QCP with a minimum in its vicinity, see Fig.~\ref{Fig:Plane}. The minima
of the isentropes indicate how the entropy accumulates around the QCP and sit
at the positions where the system is maximally undecided which ground state to
choose. According to Eq.~(\ref{PressureCaloric}), the Gr\"uneisen parameter is
proportional to the slope of the isentropes, i.e.~it has a different sign on
each side of the QCP. It thus follows from a generic entropy distribution
around the QCP that the Gr\"uneisen parameter will change sign in its
vicinity. The sign change occurs at the location of the isentrope minima and
therefore tracks the accumulation of entropy in the phase diagram. In this
way, the Gr\"uneisen parameter maps out the entropy landscape near a quantum
critical point.

Since the specific heat is always positive the sign change of the Gr\"uneisen
parameter coincides with that of the thermal expansion. As the thermal
expansion is given by the negative derivative of the entropy with respect to
pressure, $\alpha \propto -\partial S/\partial p|_{T}$, a negative thermal
expansion is present whenever the entropy increases as a function of
pressure. This happens naturally in the vicinity of a pressure-controlled
QPT. The sign change of the thermal expansion and hence of the Gr\"uneisen
parameter occurs at a pressure value where the entropy reaches a maximum.
Similarly, for magnetic field tuning the magnetocaloric effect and the
quantity $(\partial M/\partial T)_H$ will change its sign at the accumulation
point of entropy.

As mentioned above, already the scaling result for the Gr\"uneisen parameter
in the low-temperature regime, (\ref{LowTRegime}), suggested a sign change of
$\Gamma$ provided that the prefactor $G_r$ has the same sign in both low-$T$
regimes.  That this is the case is again ensured by entropic constraints. The
sign of $G_r$ is determined by the relative size of the exponents $y^\pm_0$
and $d/z$. These exponents determine the behavior of the specific heat or,
alternatively, the entropy, $S$, in the low-$T$ regime, $S \sim T^{y^\pm_0}$,
and the quantum critical regime~\cite{PRL}, $S \sim T^{d/z}$, see
Fig.~\ref{Fig:Plane}.  Using again the argument, that the competition between
two different ground states leads to an enhanced entropy close to the QCP, we
expect on physical grounds that the entropy as a function of $T$ decreases in
the low-$T$ regime at least as fast as in the quantum critical regime. This
implies that $d/z \le y^\pm_0$ and finally
\begin{align} \label{SignG_r}
G_r \le 0
\end{align}    
in both low-temperature regimes. There are examples of critical theories with
exponents $d/z = y_0$, so that $G_r=0$, e.g. an insulating Heisenberg
antiferromagnet with $z=1$ and $y_0=d$ on the ordered side of the phase
diagram or an itinerant antiferromagnet with $d=z=2$ and
$y_0=1$. In the latter example, logarithmic corrections ensure that the
entropy at the critical point is higher~\cite{PRL}.

In the low-temperature regime we can determine the critical isentropes
explicitly by using the scaling result Eq.~(\ref{LowTRegime}). We obtain for
$T \ll |r|^{\nu z}$
\begin{align} 
\left.T(r)\right|_{S} \propto |r|^{-G_r}\,.
\end{align}
The isentrope behaves as a powerlaw near the QCP with an exponent given by
$G_r$, Eq.~(\ref{UniPrefactor}). A minimum of the isentrope directly follows
if $G_r$ is negative\cite{comment}.

In the following we will give examples of two possible scenarios. If the
entropy landscape at finite temperature is only determined by the
zero-temperature transition we expect that the minima of the isentropes are
located approximately above the QCP, i.e.~at $r \approx 0$. In this scenario,
considered in Sec.~\ref{Sec:IsingChain}, the Gr\"uneisen parameter and the
magnetocaloric effect change their sign near the critical pressure $p_c$ and
critical field $H_c$.
If the QCP is however an endpoint of a line of classical finite temperature
transitions we expect that the entropy landscape is distorted with the minima
in the vicinity of the critical temperature as sketched in
Fig.~\ref{Fig:IsentropesBEC}. We will show that the sign change of the
Gr\"uneisen parameter then occurs within the Ginzburg region of the
symmetric phase. This scenario is considered in Sec.~\ref{Sec:BEC}.

\section{Sign change of $\Gamma$ near a QCP without order at finite $T$}
\label{Sec:IsingChain}

\subsection{Ising chain in a transverse field}

\begin{figure}
\centering
\includegraphics[width=0.8 \linewidth]{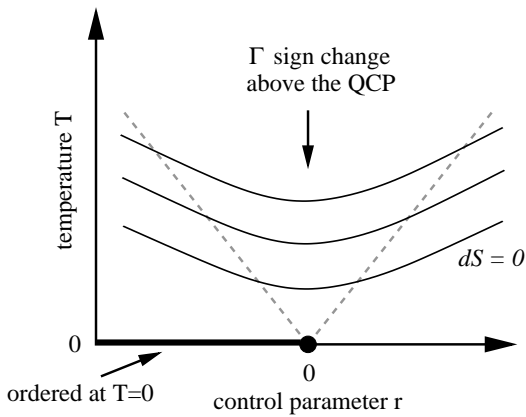}
\caption{\label{Fig:IsentropesIsing} Sketch of the isentropes near the QCP of
  the Ising chain in magnetic field with $r \propto H-H_c$. The entropy
  accumulates above the QCP leading to a sign change of the magnetocaloric
  effect (\ref{MagnetoCaloric}) at $r=0$.}
\end{figure}

\begin{figure}
\centering
\includegraphics[width=0.9 \linewidth]{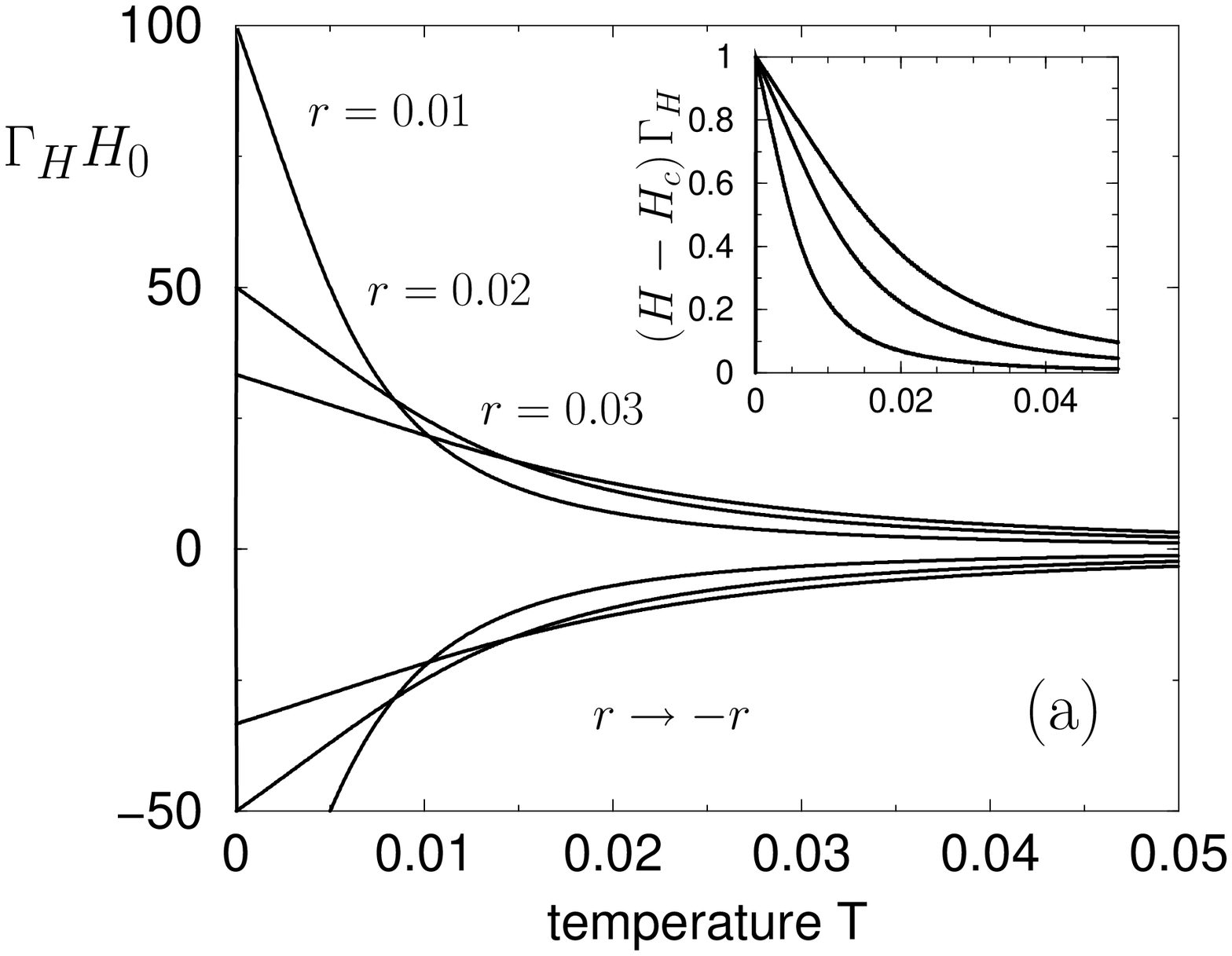}
\includegraphics[width=0.9 \linewidth]{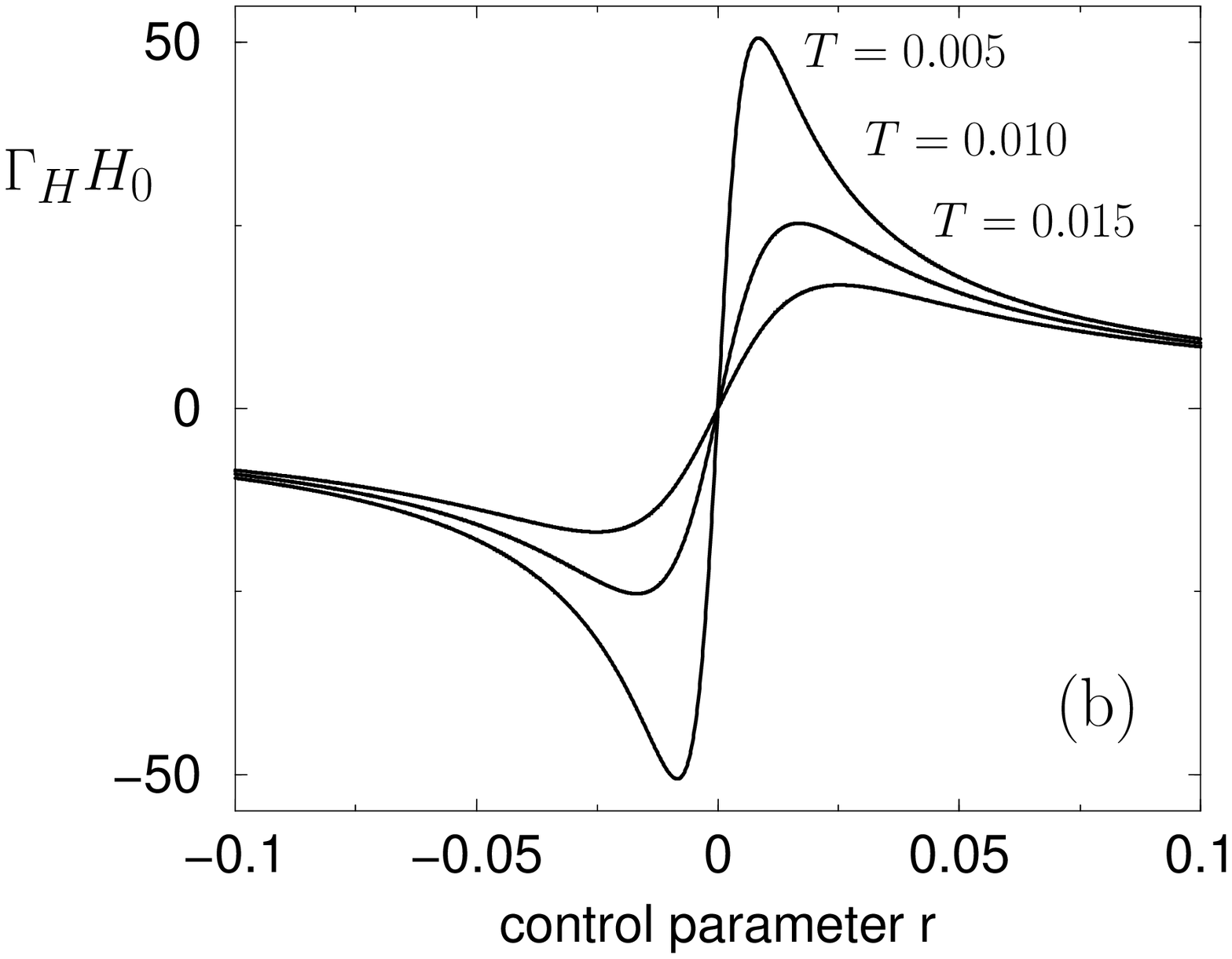}
\caption{\label{Fig:IsingMagnCalo} Magnetocaloric effect
  (\ref{MagnetoCaloric}) of the Ising chain near the critical field $H_c$ as a
  function of the control parameter $r=(H-H_c)/H_0$ for (a) a temperature
  sweep at constant magnetic field and (b) a magnetic field sweep at constant
  temperature. As shown in the inset, the saturation value in the low-$T$ regime has the universal
  prefactor, $-G_r = \nu z = 1$.}
\end{figure}

A simple example of a QPT where the entropy landscape is solely determined by
the underlying QCP is provided by the model of an Ising chain in a transverse
field. The Ising chain shows a QPT as a function of transverse magnetic field
$H$ between a magnetic and a paramagnetic ground state.  We are interested in
the behavior of the magnetocaloric effect (\ref{MagnetoCaloric}) near the
critical field $H_c$, which was also considered in
Ref.~\onlinecite{Honecker}. The continuum theory describing the transition is
given by (Majorana-) fermions with a relativistic spectrum\cite{Sachdev99}
\begin{align}
\epsilon_k = \sqrt{r^2 + k^2}\,,
\end{align}
where $r \propto H-H_c$.  The important exponents of the critical theory are
%
$z=d=\nu=1$.
%
Furthermore, the spectrum is gapped away from the QCP which leads to a
specific heat that decays exponentially with temperature. The exponent $y^\pm_0$
appearing in (\ref{UniPrefactor}) can effectively be set to infinity on both
sides of the QCP, so that the prefactor in both
low-temperature regimes simplifies to
\begin{align} \label{IsingUnivPref}
G_r = - \nu z = - 1\,.
\end{align}
The thermodynamic quantities can be computed from the free energy density
\begin{align}
f(r,T) = - T \int\limits_{-\infty}^\infty \frac{dk}{2\pi} \log 
\left[2 \cosh \frac{\epsilon_k}{2 T}\right]\,.
\end{align}
The isentropes near the critical field $H_c$ are sketched in
Fig.~\ref{Fig:IsentropesIsing}. Since the control parameter enters the free
energy only quadratically the entropy landscape is symmetric with respect to
the reflection $r \to -r$. This symmetry is rooted in the self-duality of the
theory describing the QPT\cite{privateSenthil}. This has the consequence that
in the case of the Ising chain the sign change of the magnetocaloric effect
occurs directly at $r=0$. The self-duality symmetry thus causes the prefactor
of the divergence (\ref{QCRegime}) to vanish in the quantum critical regime.
The resulting magnetocaloric effect is shown in Fig.~\ref{Fig:IsingMagnCalo}
for a temperature sweep at constant magnetic field and vice versa. In the
low-temperature regimes the magnetocaloric effect diverges with the universal
prefactor (\ref{IsingUnivPref}) as is shown in the inset of
Fig.~\ref{Fig:IsingMagnCalo}a. The developing divergence combined with a sign
change leads to a strong signature of the magnetocaloric effect in the field
sweep shown in Fig.~\ref{Fig:IsingMagnCalo}b.


\subsection{Experiments on metamagnetic quantum criticality}

An especially interesting example of a QCP without an associated finite-$T$
phase transition are the so-called ''metamagnetic quantum critical endpoints''
\cite{Millis02}. The QCP is here understood to be the endpoint, $(H_c,T^*)$,
of a line of first order transitions where the temperature $T^*$ is tuned to
zero. As the control parameter (i.e. the magnetic field) couples here linearly
to the order parameter\cite{Millis02}, the quantum-critical properties of such
systems differ in some aspects from the examples considered in this paper --
for a detailed discussion we refer to Ref.~\onlinecite{Gegenwart05}. The
notion of metamagnetic quantum criticality~\cite{Millis02} was originally
motivated by experiments\cite{Perry01,Grigera01} on Sr$_3$Ru$_2$O$_7$. The
metamagnetic transition is sensitive to pressure and magnetic field variations
suggesting that both $p$ and $H$ can be used as tuning parameters, $r=r(p,H)$,
and relations such as Eq.~(\ref{Relation-PM-Variation}) are expected to
hold. In fact, the differential susceptibility and magnetostriction nicely
track each other\cite{Grigera04} near the critical field, $H_c \approx 7.9 T$,
confirming that pressure and magnetic field variations probe indeed the same
thermodynamic information. The thermal expansion~\cite{Gegenwart05} shows a
change of sign near the critical field indicating the accumulation of entropy
in the $(H,T)$ plane above $H=H_c$.

A metamagnetic anomaly of a different type but with qualitative similar
signatures is also observed in the heavy-fermion compound
CeRu$_2$Si$_2$~\cite{Paulsen,FlouquetReview}. The thermal expansion as a
function of $H$ again shows a pronounced sign change near
the metamagnetic field $H_m \approx 7.8 T$ suggesting the vicinity to a
QCP. According to Eq.~(\ref{LowTRegime}), the zero-temperature limit of the
Gr\"uneisen parameter is expected to diverge as $\Gamma_{\rm cr}(T=0) \propto
1/(H-H_c)$ which also seems to be compatible with experimental observations,
see Fig.~16 of Ref.~\onlinecite{FlouquetReview}.

\section{Sign change of $\Gamma$ near a QCP with order at finite $T$} 
\label{Sec:BEC}

\begin{figure}
\centering
\includegraphics[width=0.9 \linewidth]{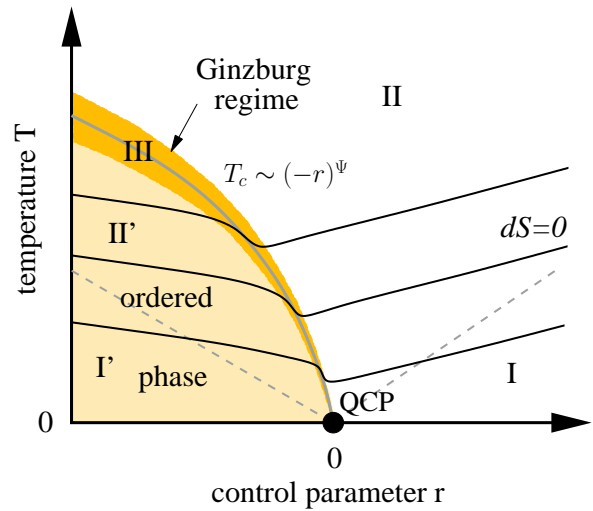}
\caption{\label{Fig:IsentropesBEC} Sketch of a phase diagram with a
  symmetry-broken phase extended to finite temperature; here the QCP is the
  endpoint of a line of classical second order transitions. The minima of the
  isentropes, which identify the position of the sign change of the
  Gr\"uneisen parameter, are located within the Ginzburg regime in the
  symmetric phase. Furthermore, in the Ginzburg regime the isentrope tend to
  nestle to the phase boundary as explained in the main text. For later
  reference the different regimes were labeled as: I/I' low-temperature
  regime, $T \ll |r|^{\nu z}$, and II/II' quantum critical regime, $T \gg
  |r|^{\nu z}$, in the symmetric/ordered phase, and III the Ginzburg regime
  that houses the phase boundary, $T_c \sim (-r)^\Psi$.  }
\end{figure}

It is a common situation that the symmetry-broken phase of the QPT extends to
finite temperature, $T$. The QCP is then in fact the endpoint of a line of
classical transitions along which derivatives of the free energy show a
singular behavior at a finite critical temperature $T_c$. Entropy is expected
to accumulate near the phase boundary and the entropy landscape is
correspondingly distorted as shown in Fig.~\ref{Fig:IsentropesBEC}.

As a consequence, the Gr\"uneisen parameter is expected to change its sign
near the critical temperature $T_c$. This is observed for example in the heavy
fermion compound CeCu$_{6-x}$Au$_x$ \cite{Estrela} and in the spin-gap
compound TlCuCl$_3$ \cite{Johannsen}. An especially nice example is the
superconducting heavy fermion system URu$_2$Si$_2$. The thermal expansion
shows a pronounced jump at the superconducting transition at $T_c = 1.18 K$
accompanied with a change of sign\cite{VanDijk} suggesting the vicinity of an
associated QCP. The superconducting condensate of URu$_2$Si$_2$ forms in fact
within a so-far unidentified ``hidden order'' phase which can be suppressed by
an applied magnetic field of $H_c \approx 35.9 T$~\cite{Jaime02}. Moreover, an
additional reentrant phase is located in a magnetic field range of $H = 36 -
39 T$. The magnetocaloric effect has been measured~\cite{Jaime02} and the
isentropes around the second order transition at $H\approx 36 T$ have a
similar shape as in Fig.~\ref{Fig:IsentropesBEC} with minima
near the critical temperature, $T_c$. Another example is the heavy-fermion
alloy U(Pt,Pd)$_3$ where a Gr\"uneisen parameter inversion was also
observed \cite{deVisser}.

As an illustration of such a scenario we will consider the quantum phase
transition of the dilute Bose gas. This example will capture many features
which we believe are generic for a QCP with order at finite temperature. We
assume that the chemical potential is sensitive to pressure so that a quantum
phase transition can be induced by tuning the pressure to a critical value
$p_c$. Moreover, for dimensions $d>2$ a finite temperature transition is
present which distorts the entropy landscape in a manner as sketched in
Fig.~\ref{Fig:IsentropesBEC}. The dilute Bose gas is relevant for the
field-driven QPT in the spin-gap compounds which is interpreted as the Bose
condensation of magnons~\cite{Giamarchi,Rueegg,Johannsen,Jaime04,Sebastian,Inga05}.

The action of the dilute Bose gas has the form\cite{Sachdev99}
\begin{align} \label{BEC-action}
\mathcal{S}[\phi,\phi^*] =
\int_0^\beta\! d\tau \int d^d\br
& \left[\phi^*(\tau,\br) 
\left(\frac{\partial}{\partial \tau}-\nabla^2-\mu_0\right) 
\phi(\tau,\br)
\right.\nn
\\
&\left.\quad
+ \frac{u}{2}\, |\phi(\tau,\br)|^4\right]\,.
\end{align}
We will limit ourselves to a discussion of this model above the upper critical
dimension of the QPT, $d+z>4$ and therefore $d>2$, where the correlation length and the dynamical exponent of
the QPT are given by their Gaussian values,
\begin{equation}
\nu = \frac{1}{2},\quad z=2\,.
\end{equation}

First let us consider the theory on the mean-field level, however, {\em
including} the one-loop correction to the chemical potential. The Landau
potential takes the form
\begin{align} \label{MeanField}
\mathcal{V}_{\rm MF}(\phi, \phi^*) &= R |\phi|^2 + \frac{u}{2} |\phi|^4\,.
\end{align}
where the effective mass $R$ is temperature dependent as it is renormalized by
the critical fluctuations,
\begin{align} \label{EffMass}
R = -\mu_0 + 2 u \int \frac{d^d\bk}{(2\pi)^d} 
\frac{1}{2} \coth\left(\frac{\bk^2}{2 T}\right)
=
r -  r_{\rm cr}(T)\,.
\end{align}
In the last equation we introduced the control parameter $r$ given in terms of
the renormalized zero-temperature chemical potential, $r \equiv - \mu = -
\mu_0 + u \int \frac{d^d\bk}{(2\pi)^d}$. If the QCP is sensitive to pressure
changes, $r$ can be linearized in the applied pressure, $r = (p-p_c)/p_0$
(assuming a finite magnetic field in the case of critical spin-gap
compounds\cite{Inga05}), see Eq.~(\ref{ControlParameter}). The phase
transition occurs when the mass vanishes, $R=0$.  The phase boundary in the
plane $(r, T)$ is thus given by\cite{Sachdev99}
\begin{equation} \label{PhaseBoundary}
r_{\rm cr}(T) = -\frac{\zeta(d/2)}{2^{d-1} \pi^{d/2}} u\,T^{1/\psi}\,.
\end{equation}
where the exponent reads $1/\psi = (d+z-2)/z = d/2$.  Note that the
temperature dependence of the phase boundary, $T_c \propto
(p_c-p)^{\psi}$, is due to the dangerously irrelevant quartic coupling $u$.

Above the upper critical dimension, $d+z > 4$, irrelevant couplings will in
general invalidate the simple scaling form of the critical free energy density
given in Eq.~(\ref{ScalingForm}). In the present example the dependence on the
quartic coupling, $u$, must be incorporated into a generalized scaling form
\begin{align} \label{ScalingFormWithu}
f(r,T,u) = b^{-(d+z)} f(r\, b^{1/\nu},T\, b^z, u\, b^{d+z-4})\,.
\end{align}
The dependence on $u$ may spoil the predictions for the Gr\"uneisen parameter
drawn from Eq.~(\ref{ScalingForm}). This is for example the case within the
wedge of the Ginzburg region indicated in Fig.~\ref{Fig:IsentropesBEC} which
contains the phase boundary. Bearing this in mind we now go beyond the mean
field treatment by including the contribution of fluctuations in the free
energy density.

\subsection{Gaussian approximation}

Depending on the regime in the phase-diagram plane,
Fig.~\ref{Fig:IsentropesBEC}, the physics of the dilute Bose gas is determined
by different fixed points\cite{Weichman}. 

In the low-temperature regime I' within the symmetry-broken phase, $T \lesssim
-r$, the physics of the Gr\"uneisen parameter 
will be controlled by the Goldstone modes; an expansion around the
Gaussian theory in the quartic coupling $u$ is plagued with IR divergencies in
this regime indicating the crossover to the stable Goldstone fixed
point\cite{Weichman,Pistolesi,Schakel}.  This regime is conveniently treated
by decomposing the fluctuations into massive amplitude- and gapless phase
modes.  The thermodynamics will be determined by the phase modes leading to a
specific heat which decreases in temperature as $C_p \sim (T/\sqrt{-r})^{y^-_0}$ with
$y^-_0 = d$. The prefactor, $G_r$, of the Gr\"uneisen divergence
(\ref{LowTRegime}) is here given by
\begin{align} \label{PrefactorGoldstoneRegime}
\left.-G_r\right|_{-} = -\frac{\nu(d - z y^-_0)}{y^-_0} = \frac{1}{2}\,.
\end{align}

On the other hand, the thermodynamics in the low-temperature regime I within the
symmetric phase, $T \lesssim r$, is determined by the Gaussian fluctuations
around the $T=0$ theory. The specific heat decreases exponentially with
temperature, $C_p \sim r^{d/2} (r/T)^{1/2} e^{-r/T}$, reflecting the gap in
the fluctuation spectrum. Correspondingly, in this regime the prefactor $G_r$ is
given by
\begin{align}
\left.-G_r\right|_{+} = \nu z = 1\,.
\end{align}

The most interesting regime for our discussion is the quantum critical regime,
$T \gg |r|$. Here thermodynamics is almost always (in regions indicated as II
and II') dominated by ideal gas behavior with a specific heat, $C_p \sim
T^{d/2}$ and thermal expansion $\alpha \sim T^{d/2-1}$. Only sufficiently near
the classical critical transition in regime III, thermodynamics will be
controlled by the classical critical fixed point which for the dilute Bose gas
belongs to the XY-universality class. This crossover occurs at the Ginzburg
temperature when the Ginzburg parameter is of order one,
\begin{align} \label{Ginzburg}
\mathcal{G} \equiv u\, T |R|^{(d-4)/2} = \mathcal{O}(1)\,.
\end{align}
Note that the effective classical quartic coupling is given by $u\, T$. The
description of the classical critical properties within this Ginzburg regime
is beyond the simple approximation scheme employed here. It is within the
Ginzburg regime where the thermal expansion and the Gr\"uneisen parameter
change sign. However, as we will explain in detail below the sign change is
not a property associated to classical criticality but is rather to be
attributed to the underlying quantum phase transition. The sign change is a
property of the quantum critical background on which the classical
singularities develop.

For its description we will evoke a Gaussian approximation which captures the
correct thermodynamics in the quantum critical regime, $T \gg |r|$, except in
the Ginzburg region where it will fail yielding singularities with wrong
Gaussian exponents. In the quantum critical regime within the symmetric phase the Gaussian
fluctuations determine the free energy density
\begin{align} \label{FreeEnergySymm}
f_+ =  
T \int \frac{d^d\bk}{(2\pi)^d} 
\log \left[1 - \exp\left(- \frac{R(r,T) + \bk^2}{T}\right)\right]\,.
\end{align}
For the following it will be crucial that here and in
Eq.~(\ref{FreeEnergyBroken}) the temperature dependent mass $R(r,T)$ enters,
see Eq.~(\ref{EffMass}), which measures the distance from the classical
transition.  Note that in Eq.~(\ref{EffMass}) we have set $R=0$ on the
right-hand side, using only critical fluctuations when computing the mass
renormalization.  For our discussion it is essential that this approximation
is justified outside of the Ginzburg regime, as corrections are of order of
the Ginzburg parameter, $\delta R/R = \mathcal{O}(\mathcal{G})$. Outside the
Ginzburg regime, we therefore obtain similar results as other approximation
schemes like the self-consistent Hartree-Fock or the Popov
approximation. Those approximations have, however, the disadvantage that they
wrongly predict a first-order transition~\cite{Baym,Andersen} within the
Ginzburg regime, while our approximation describes a second-order phase
transition with Gaussian (and therefore wrong) critical exponents.

In the symmetry-broken phase the field fluctuates around the solution of the
mean-field potential (\ref{MeanField}). The condensate then attains the finite
value, $|\phi|^2 = -R/u$, and contributes to the free energy. The fluctuations
around the finite condensate have the Bogoliubov spectrum,
\begin{align} \label{FreeEnergyBroken}
f_- &= \frac{R (R - 2 r)}{2 u} 
\\\nn & + T \int \frac{d^d\bk}{(2\pi)^d} 
\log \left[1 -\exp\left(- \frac{\sqrt{\bk^2
\left(-2 R + \bk^2\right)}}{T}\right)
\right]\,.
\end{align}
Note that by virtue of (\ref{EffMass}) the resulting entropy derived within
this Gaussian approximation is indeed continuous at the critical temperature
as is appropriate for a second order phase transition.

The temperature dependence of the free energies, (\ref{FreeEnergySymm}) and
(\ref{FreeEnergyBroken}), is two-fold: there is an explicit $T$-dependence and
an implicit dependence via the effective mass $R = R(r,T)$. In the following
it will be useful to distinguish between a quantum critical and a classical
critical contribution to thermodynamics. We will define the quantum critical
contribution to be the one which derives from the explicit temperature
dependence of the free energies. The classical contribution to thermodynamics
will arise from the implicit $T$-dependence via the effective mass $R(r,T)$
which is induced by the dangerously irrelevant quartic coupling $u$, see
Eq.~(\ref{PhaseBoundary}). This latter contribution will dominate
thermodynamics near the classical transition since near $T_c$ the free energy
is very sensitive to variation of $R$.

To illustrate this point let us rewrite the free energy densities
(\ref{FreeEnergySymm}) and (\ref{FreeEnergyBroken}),
\begin{align} \label{FreeEnergy-Scaling}
f_\pm  =  \Psi_\pm(r, R) 
+ T^{(d+z)/z} \mathcal{F}_\pm( R \,T^{-1/(\nu z)})\,,
\end{align}
where $\Psi_+ = 0$ and $\Psi_-(r, R) = (R (R - 2 r))/(2 u)$ is the
contribution from the condensate. The part of the free energies that depends
explicitly on temperature obeys scaling with $z = 1/\nu = 2$ and the scaling
functions
\begin{align} \label{ScalingFunctions}
\mathcal{F}_\pm(x) &= \frac{K_d}{2} \int\limits_0^\infty d t\, t^{(d-2)/2} 
\log\left[1 - e^{- \omega_\pm(t,x)}\right] 
\end{align}
where $K^{-1}_d = 2^{d-1} \pi^{d/2} \Gamma(d/2)$ and 
\begin{align}
\omega_+(t,x) &= t + x\,,
\qquad
\omega_-(t,x) = \sqrt{t(t - 2 x)}\,.
\end{align}
The Gaussian approximation yields an expression for the free energy density
which conforms to the general scaling form (\ref{ScalingFormWithu}). The
dependence on the quartic coupling, $u$, however appears only via the thermal
renormalization of the mass $R = R(r,T)$.  Apart from this implicit
temperature dependence induced by the dangerously irrelevant quartic coupling,
$u$, the expression (\ref{FreeEnergy-Scaling}) resembles the quantum critical
scaling form (\ref{ScalingForm}).  Thermodynamic contributions that only
derive from the explicit temperature dependence will therefore conform with
the results obtained from Eq.~(\ref{ScalingForm}). In this sense it is
appropriate to call these contributions quantum critical. The implicit
temperature dependence via $R=R(r,T)$ results in additional classical
contributions to thermodynamics which are subleading except in the Ginzburg
regime where the Gaussian approximation breaks down. The purpose of including
the thermal renormalization of the mass is to ensure the correct threshold
behavior at $T_c$. The sign change of $R(r,T)$ across the phase transition
will reflect itself in a sign change of the Gr\"uneisen parameter. This sign
change persists even outside the Ginzburg region as is explained below.

\subsection{Sign change of the Gr\"uneisen parameter}

We will show in the following that the sign change of the Gr\"uneisen
parameter stems from the quantum critical contribution to the thermal
expansion and is in fact a property of the functions
$\mathcal{F}_\pm$. Consider the contribution to the thermal expansion at the
critical temperature $T_c$ deriving from the explicit temperature dependence,
\begin{align}
\alpha^\pm_{\rm QCP} 
&\equiv 
\frac{1}{V_m p_0}\, \underset{R \to 0}{\rm lim} 
\,\frac{\partial^2}{\partial R \partial T} f_\pm 
\\\nn
&= 
\frac{1}{V_m p_0} \frac{\nu(d+z)-1}{\nu z} T_c^{\frac{\nu d-1}{\nu z}} 
\mathcal{F}'_\pm(0)\,.
\end{align}
Although the scaling function itself is continuous at the phase transition,
$\mathcal{F}_+(0) = \mathcal{F}_-(0)$, its derivative is not. In our specific
example we have 
\begin{align}
\mathcal{F}'_+(0) = - \mathcal{F}'_-(0)
\,,
\end{align}
i.e.~the quantum critical contribution to the thermal expansion is
discontinuous and changes sign at the phase transition. Note that the
quantum critical contribution to the specific heat is smooth at
$T_c$.

Making use of Eq.~(\ref{EffMass}), the associated anomaly in the
thermal expansion, $\Delta\alpha_{\rm QCP} \equiv \alpha^+_{\rm QCP} -
\alpha^-_{\rm QCP}$, can be related to the derivative of the phase boundary,
\begin{align} \label{AnomalyThermExpQCP}
\Delta \alpha_{\rm QCP}
= - \frac{1}{V_m p_0} \frac{1}{u} 
\left.\frac{\partial r_{\rm cr}(T)}{\partial T}\right|_{R=0}
= - \frac{1}{V_m p^2_0 u} \left[\frac{d T_c}{d p}\right]^{-1}
\,.
\end{align}
The same anomaly follows also from the renormalized mean-field potential
(\ref{MeanField}). In this sense the anomaly $\Delta\alpha_{\rm QCP}$ and the
resulting sudden sign change of the Gr\"uneisen parameter near $T_c$ can be
interpreted as a finite temperature manifestation of the mean-field character
of the underlying quantum phase transition. Although the sharp jump will be
smeared out by the classical critical fluctuations in spatial dimensions $d <
4$, the smearing is confined to the Ginzburg regime which is vanishingly small
near the QCP. A pronounced jump near $T_c$ in the thermal expansion and in the
Gr\"uneisen parameter with an accompanying sign change will result.

The smearing of the jump will also shift the exact position of the sign change
away from the critical temperature $T_c$. In the following we will argue that
this shift is towards the symmetric phase.

\subsection{Location of the sign change}

Where is the position of the sign change and hence the minima of the
isentropes exactly located? Let us first give some general arguments.  Let us
consider the behavior of the entropy upon approaching the phase boundary from
the ordered phase, see Fig.~\ref{Fig:IsentropesBEC}. The entropy attributed to
the QCP increases when the phase boundary is approached from the ordered side
by increasing the control parameter, $\partial S/\partial r|_{T<T_c} >
0$. When we enter the Ginzburg regime the change in entropy becomes dominated
by the finite temperature phase transition. The symmetric phase can be entered
by either increasing control parameter $r$ or temperature $T$. However, within
the Ginzburg regime  the tuning of $r$, i.e., pressure or
temperature have the same effect since both parameters couple to the same
relevant operator of the classical transition.
Since the entropy always increases as a function of temperature we also have
$\partial S/\partial r|_{T=T_c} > 0$. The entropy as a function of $r$ should
therefore attain its maximum, $\partial S/\partial r = 0$, above the critical
temperature $T_c$. As a consequence, it follows that the sign change of the
thermal expansion and the Gr\"uneisen parameter should occur within the
symmetric phase.

We can obtain an explicit expression for the Gr\"uneisen parameter at the
critical temperature, $T_c$, when the specific heat is sufficiently singular
at the classical second order phase transition. The derivation follows
standard arguments\cite{Landau}. Near the finite temperature
transition the singular part of the molar entropy can be written in the form
\begin{align}
S_{\rm CL} = S_{\rm CL}(T-T_c(p),p)\,.
\end{align}   
Near the critical temperature, the leading contribution to the thermal
expansion will derive from the pressure dependence of the first argument. We
thus obtain for the critical thermal expansion
\begin{align}
V_m \alpha_{\rm cr} \sim 
- \left.\frac{\partial S_{\rm CL}}{\partial p}\right|_T
\sim \frac{\partial S_{\rm CL}}{\partial T} \frac{d T_c}{dp}
\sim \frac{C_{\rm cr}}{T_c} \frac{d T_c}{dp}\,.
\end{align}
If the classical critical contribution is sufficiently singular such that the
background contribution can be neglected the Gr\"uneisen parameter at $T_c$
is just given by the slope of the phase boundary,
\begin{align} \label{ClassicalGamma}
\Gamma(T = T_c) = \frac{1}{V_m T_c} \frac{d T_c}{d p} 
= - \left.\frac{\Psi}{V_m (p_c-p)}\right|_{T=T_c(p)}\,.
\end{align}
In the last equation we made use of $T_c \propto (p_c-p)^\Psi$.  The negative
slope of the phase boundary, i.e.~the suppression of $T_c$ with increasing
pressure, $r = (p-p_c)/p_0$, as depicted in Fig.~\ref{Fig:IsentropesBEC},
results in a negative Gr\"uneisen parameter at the critical temperature,
$\Gamma(T=T_c) < 0$. The Gr\"uneisen parameter thus has the same sign at the
critical temperature as in the ordered phase. We again find that the sign
change must occur within the symmetric phase.

\begin{figure}
\centering
\psfrag{temperature T}{\hspace{-1em} temperature $T$}
\psfrag{thermal}{$\alpha\, V_m p_c$}
\psfrag{specific}{\; $\gamma$}
\psfrag{Grueneisen}{\hspace{-5em} Gr\"uneisen parameter $\Gamma\, V_m p_c$}
\includegraphics[width= 0.75\linewidth,angle=-90]{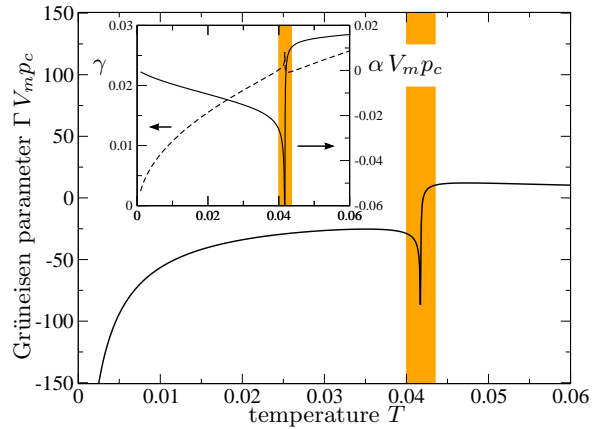}
\caption{\label{Fig:BECFigures} Gr\"uneisen parameter of the dilute Bose gas
as a function of temperature in dimension $d=3$ as derived from the Gaussian
approximation to the free energy, (\ref{FreeEnergySymm}) and
(\ref{FreeEnergyBroken}). The sign change occurs near the classical
transition. The shaded area indicate the Ginzburg regime where the employed
Gaussian approximation breaks down. The inset shows the temperature evolution
of the specific heat coefficient and the thermal expansion. The value of the
control parameter was chosen $r=-0.001$, temperature is shown in dimensionless
units and the quartic coupling has been set to $u=1$.}
\end{figure}

In order to locate the sign change  
we consider the thermal expansion of the dilute Bose gas within
the symmetric phase. The quantum critical contribution reads
\begin{align}
\alpha_{\rm QCP} &= 
\frac{1}{V_m p_0} \frac{\partial^2 f_+(R(r,T),T)}{\partial R \partial T} 
\nn\\&=
\frac{1}{2 u V_m p_0} \frac{\partial R}{\partial T} 
\left(1 +  \mathcal{O}( u T R^{(d-4)/2} ) \right)
\end{align}
In addition to $\alpha_{\rm QCP}$, the implicit temperature dependence via the
effective mass, $R$, yields a classical contribution
\begin{align}
\alpha_{\rm CL} = \frac{1}{V_m p_0} \frac{\partial^2 f_+(R(r,T),T)}{\partial
R^2} \frac{\partial R}{\partial T} \,.
\end{align}
We obtain for their ratio
\begin{align}
0 > \frac{\alpha_{\rm CL}}{\alpha_{\rm QCP}} \sim 
2 u \frac{\partial^2 f_+}{\partial R^2}
= \mathcal{O}( u T R^{(d-4)/2} )\,.
\end{align}
The classical part of the thermal expansion has a sign opposite to the part
attributed to the QCP. The classical part therefore reduces the contribution
$\alpha_{\rm QCP}$ upon approaching the phase boundary from the symmetric
phase. This finally leads to the sought-after sign change.  The two
contributions are of the same order when the Ginzburg criterion
(\ref{Ginzburg}) is fulfilled. The position of the sign change of the
Gr\"uneisen parameter is hence located within the Ginzburg regime of the
classical finite temperature transition. Sufficiently near the QCP the
Ginzburg regime is vanishingly small so that the position of the sign change
of $\Gamma$ almost coincides with the critical temperature $T_c(p)$.

The temperature evolution of the specific heat, thermal expansion and the
Gr\"uneisen parameter is shown in Fig.~\ref{Fig:BECFigures}. The control
parameter has been chosen negative such that the critical temperature is
crossed during the temperature sweep. The temperature region shown in
Fig.~\ref{Fig:BECFigures} corresponds to the quantum critical regime, $|r| \ll
T^{1/(\nu z)}$, where the expressions (\ref{FreeEnergySymm}) and
(\ref{FreeEnergyBroken}) for the free energy are applicable. In this regime
the Gr\"uneisen parameter is determined by temperature and obeys the scaling
form (\ref{QCRegime}). Since we have $\nu z = 1$ it behaves as $\Gamma \propto
1/T$. The curve follows this behavior except within the small Ginzburg regime
where the Gr\"uneisen parameter strikingly changes its sign and forms a sharp
peak at the critical temperature. Since within our Gaussian approximation the
classical critical specific heat is divergent with the Gaussian exponent,
$\alpha_{\rm Gauss} = 2-d/2 > 0$, the peak value is given by
Eq.~(\ref{ClassicalGamma}), where for the dilute Bose gas we have $\Psi =
2/d$. In particular, at the phase transition the Gr\"uneisen parameter is now
determined by the distance to the QCP, $p-p_c$. Since the phase boundary is
located well inside the quantum critical regime, $T \gg |p-p_c|$, where
$\Gamma$ is usually determined by temperature, $\Gamma \propto 1/T$, the
Gr\"uneisen parameter is strongly enhanced at $T_c$. The crossover to this
enhanced value occurs within the narrow Ginzburg regime leading to a peak
structure at $T_c$. The peak of $\Gamma$ at the critical temperature manifests
itself in an additional tilt of the isentropes within the Ginzburg regime as
sketched in Fig.~\ref{Fig:IsentropesBEC}. Indeed, remembering that the
Gr\"uneisen parameter just measures the slope of the isentropes,
Eq.~(\ref{PressureCaloric}), it follows from expression (\ref{ClassicalGamma})
that the isentrope locally follows the phase boundary at $T_c$.

The well-pronounced peak of the Gr\"uneisen parameter at $T_c$ is only
expected for a sufficiently singular classical critical specific heat such
that relation (\ref{ClassicalGamma}) holds. The Gaussian approximation
employed here overestimates the classical critical specific heat exponent of
the dilute Bose gas. In fact, the specific heat exponent of the $d=3$
XY-universality class is negative, i.e.~the specific heat is not divergent at
$T_c$. Nevertheless, we still expect it to dominate over the quantum critical
background such that a narrow peak in $\Gamma$ should evolve at $T_c$.

As can be seen in the inset of Fig.~\ref{Fig:BECFigures}, near the QCP the
anomaly in the specific heat, $\Delta C_p$, near $T_c$ is small in comparison
with the one in the thermal expansion, $\Delta \alpha$. This is expected since
the large anomaly in the thermal expansion is attributed to the quantum
critical point, $\Delta \alpha = \Delta \alpha_{\rm QCP}$, see
Eq.~(\ref{AnomalyThermExpQCP}), whereas the anomaly of the specific heat
originates only from the classical contribution to thermodynamics due to the
temperature dependence induced in the effective mass, $R$, by the dangerously
irrelevant quartic coupling, $u$.  

This is also in agreement with the Ehrenfest relation\cite{Landau} which
compares the size of the anomalies, i.e., the jumps in thermal expansion and
specific heat at the classical transition that derive from the mean-field
potential (\ref{MeanField}),
\begin{align}
\frac{\Delta \alpha}{\Delta C_p} = \frac{1}{V_m} \frac{d \log T_c(p)}{d p} =
\left.\frac{\psi}{V_m (p-p_c)}\right|_{T=T_c(p)} < 0\,.
\end{align}
In the second equation we made again use of Eq.~(\ref{PhaseBoundary}). As the
quantum critical point is approached the relative size of the anomalies is
expected to diverge as $\propto 1/(p-p_c)$ resulting in a dominant anomaly in
the thermal expansion. Note that although it has the same functional form as
Eq.~(\ref{ClassicalGamma}) the Ehrenfest relation contains different
information. In particular, the Ehrenfest relation does not discuss the
absolute size of the Gr\"uneisen parameter $\Gamma$.

\begin{figure}[t]
\centering
\includegraphics[width=0.9 \linewidth]{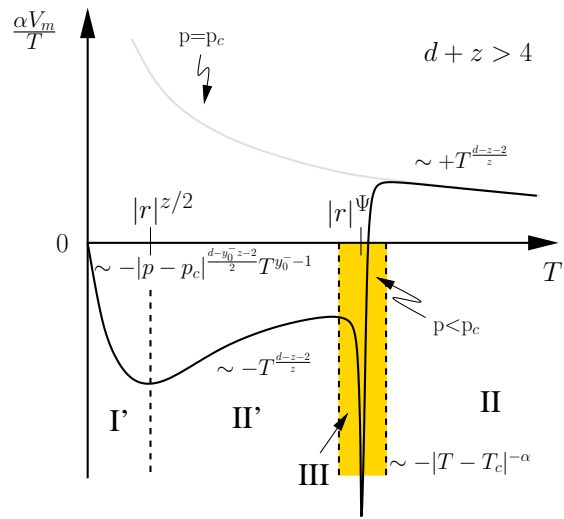}
\caption{\label{Fig:SketchThermExp} Sketch of the thermal expansion
  coefficient, $\alpha/T$, in different regimes of the phase diagram in
  Fig.~\ref{Fig:IsentropesBEC} for a QPT above its upper critical dimension,
  $d+z > 4$. The thermal expansion changes sign near the critical temperature
  $T_c$. The exponent $\alpha$ is the specific heat exponent of the classical
  transition and $y_0^-$ is determined by the spectrum of low-lying excitations
  in regime I'.}
\end{figure}

At low temperatures the Gr\"uneisen parameter will eventually saturate after
crossing over into the low-temperature regime I' (not shown) and converges to
a value now given by Eq.~(\ref{LowTRegime}). The behavior of the thermal
expansion and the Gr\"uneisen parameter in the various regimes of the phase
diagram are summarized in Fig.~\ref{Fig:SketchThermExp} and
\ref{Fig:SketchGrueneisen}.

\section{Summary}\label{Sec:Summary}

The Gr\"uneisen parameter, $\Gamma$, and the magnetocaloric effect change
sign near generic quantum critical points. We showed that the position of the
sign change indicates the accumulation of entropy in the phase diagram. Two
scenarios have been distinguished: a QCP with and without a symmetry broken
phase at finite temperature. For the latter, treated in Section
\ref{Sec:IsingChain}, the sign change is expected to be located
near the critical value of the control parameter, e.g., near the critical
pressure $p_c$. We discussed two examples where such a scenario is realized:
the Ising chain in a transverse field and metamagnetic quantum critical
materials.

\begin{figure}[t]
\centering
\includegraphics[width=0.9 \linewidth]{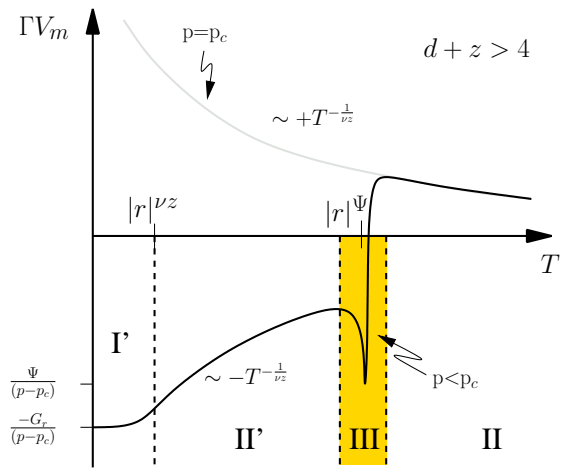}
\caption{\label{Fig:SketchGrueneisen} Sketch of the Gr\"uneisen parameter,
  $\Gamma$, (or equivalently of the magnetocaloric effect in the case of a
  field-driven transition) in different regimes of the phase diagram in
  Fig.~\ref{Fig:IsentropesBEC} for a pressure-tuned QPT, $r \propto p-p_c$,
  above its upper critical dimension, $d+z > 4$, where $\nu = 1/2$ and $\Psi =
  z/(d+z-2)$; $G_r$ is defined in Eq.~\ref{UniPrefactor}. Near the classical
  critical transition $\Gamma$ changes its sign in a characteristic jump. The
  peak at the critical temperature, $T_c \propto (p_c-p)^\psi$, is present if
  the specific heat is sufficiently singular at $T_c$, see
  Eq.~(\ref{ClassicalGamma}). The intermediate regime II' vanishes in the
  limit $d+z \to 4+$.}
\end{figure}

An overview of our main results for the second scenario, where a
symmetry-broken phase at finite $T$ is present, is given in
Figs.~\ref{Fig:SketchThermExp} and \ref{Fig:SketchGrueneisen}. While explicit
calculations have been performed for a QPT of the dilute Bose gas, we expect
that all of our qualitative results are equally valid for other quantum phase
transitions where the control parameter couples quadratically to the order
parameter. Note that the exponents for $\Gamma$ in
Fig.~\ref{Fig:SketchGrueneisen} were obtained by considering the ratio of the
critical parts of thermal expansion and specific heat; for comparison with
experiments a possible non-critical background contribution might have to be
subtracted, cf.~discussion in Ref.~\onlinecite{PRL}. Most interesting is a
situation where one investigates the behavior on the ordered side of the phase
diagram close to the QCP. Upon lowering temperature one crosses according to
Fig.~\ref{Fig:IsentropesBEC} four different regimes.

In regime II one observes the usual power-laws in thermal expansion and
Gr\"uneisen parameter (or equivalently magnetocaloric effect) which are
associated to the quantum-critical part of the phase diagram.  The main result
of our paper is the characteristic jump of thermal expansion and Gr\"uneisen
parameter in the Ginzburg regime located slightly above the classical phase
transition. The jump in $\Gamma$ proportional to $\frac{1}{(p_c-p)^{\Psi/(\nu
z)}}$, where $\Psi$ characterizes the form of the phase boundary, $T_c \propto
(p_c-p)^\Psi$, gets more and more pronounced upon approaching the QCP. If the
classical specific heat is diverging, there is also a sharp maximum in
$\Gamma$ at $T_c$ with the universal value $\Gamma(T_c)=\Psi/(V_m(p-p_c))$.
In situations where the classical specific heat is not so singular, a peak can
still occur but a lower maximal value is expected.  In the ordered phase
outside of the Ginzburg regime, i.e. in regime II', the Gr\"uneisen parameter
increases again as in II but with an opposite sign as the system is located on
the other side of the phase transition.  Finally, at lowest temperatures a
saturation of $\Gamma$ sets in to an universal value given by
Eq.~(\ref{LowTRegime}).  For a magnetic-field tuned quantum phase transition
an analogous behavior is expected for the magnetocaloric effect
(\ref{MagnetoCaloric}).  In cases where the QCP is below its upper critical
dimension, one has $\Psi=\nu z$ and region II' is absent.

It is an interesting open question how the above results are modified for
quantum phase transitions which involve different types of fluctuations
possibly characterized by different time-scales as for example in the case of
itinerant ferromagnetism\cite{Hertz,Belitz}.

To summarize, the divergence of the Gr\"uneisen parameter and magnetocaloric
effect~\cite{PRL} near a QCP in combination with their sign change result in
very strong signatures. They are thus important thermodynamic probes to
detect and classify QCPs.

\acknowledgments

The authors acknowledge fruitful discussions with P.~Gegenwart and K.~Grube
and a collaboration with L. Zhu and Q. Si at an earlier stage of this work.
This work was supported by the Deutsche Forschungsgemeinschaft under grant GA
1072/1-1 and  the SFB 608.


\begin{thebibliography}{}
\bibitem{comment} 
We use the notation of Ref.~\onlinecite{PRL} where the
factor $G_r$ was introduced.  The chosen sign convention is unfortunate.
\end{thebibliography}
\end{document}